\documentclass[aps,pra,showpacs,twocolumn,superscriptaddress]{revtex4-1}
\usepackage{amssymb}
\usepackage{graphicx}
\usepackage{appendix}
\usepackage{dcolumn}
\usepackage{bm}
\usepackage{amsmath}
\usepackage{epstopdf}
\usepackage{xcolor}
\usepackage{float}
\usepackage{braket}
\usepackage[english]{babel}
\usepackage{epstopdf}
\usepackage[colorlinks,linkcolor=magenta,citecolor=blue,urlcolor=blue]{hyperref}

\begin{document}

\title{Majorana zero modes, unconventional real-complex transition and mobility edges in a one-dimensional non-Hermitian quasi-periodic lattice}
\author{Shujie Cheng}
\affiliation{Department of Physics, Zhejiang Normal University, Jinhua 321004, China}

\author{Gao Xianlong}
\thanks{gaoxl@zjnu.edu.cn}
\affiliation{Department of Physics, Zhejiang Normal University, Jinhua 321004, China}

\date{\today}

\begin{abstract}
 In this paper, a one-dimensional non-Hermitian quasiperiodic $p$-wave superconductor without $\mathcal{PT}$-symmetry is studied. By analyzing the spectrum, we 
 discovered there still exists real-complex energy transition even if the inexistence of $\mathcal{PT}$-symmetry breaking. By the inverse participation ratio, we constructed such 
 a correspondence that pure real energies correspond to the extended states and complex energies correspond to the localized states, and this correspondence is 
 precise and effective to detect the mobility edges. After investigating the topological properties, we arrive at a fact that the Majorana zero modes in this system are immune 
 to the non-Hermiticity.  
\end{abstract}

\maketitle

\section{Introduction}
In 1958, P. W. Anderson uncovered that the absence of the diffusion of wave packets is induced by disorder \cite{Anderson}. Since then, Anderson localization 
has gradually appealed much attention and become an active research area in condensed-matter physics.  The scaling theory tells us that when the strength of 
the disorder reaches the threshold, all the eigenstates of the one-dimensional (1D) and 2D systems will be the Anderson localized states. However, Mott found that 
in some exceptional systems, i.e. the 3D Anderson model, only part of the eigenstates are localized, which are separated from the extended states by the 
so-called mobility edges \cite{Mott}. 

In reality, beyond the 3D systems, in the 1D quasi-periodic systems, known as the Aubry-Andr\'{e}-like (AA-like) models (extensions of AA model \cite{AA}), 
there also occurs mobility edges (MEs). Here, the quasi-periodicity accounts for the uncorrelated disorder. In 1988, Sarma et.al. discovered that in a 
class of AA-like model with slow-varying potentials, there also existes MEs \cite{Sarma_1}. The numerical solutions show that the density of states peak at MEs, 
and the first-order derivative of Lyapunov exponent is discontinuous at MEs. Later, a asymptotic semiclassical technique is proposed to locate the MEs \cite{Sarma_2}. 
The authors transformed the problem of solving the mobility edge into the problem of analyzing the solution of a characteristic equation. The complex solutions correspond 
to the extended states, and the real solutions correspond to the localized states. Accordingly, mobility edges are acquired by this strategy. Recently, this analytical 
method is developed to calculate the MEs in off-diagonal AA-like models. When the slow-varying potential is incommensurate and the hopping amplitudes are 
commensurate, Liu et.al. found that there are two pairs of MEs at weak potential strength, and the system becomes Anderson localized when the potential gets stronger \cite{Tong_1}. 
When the off-diagonal term is incommensurate and the potential is commensurate, there is a pair of parallel MEs; when the two terms are both incommensurate, 
there will appear multiple MEs, and the singularity at which the MEs intersect signals the Anderson localization \cite{Tong_2}. Furthermore, Liu et.al. 
investigated the delocalization-localization properties of another form of off-diagonal AA-like model \cite{Tong_2_1}. The results show that the wave functions present multifractal 
behavior, thus making the phase diagram consist of extended phase and critically localized phase. Recently, this theoretically model has been experimentally realized by  
taking advantage of ultracold atomic momentum-lattice engineering  and these phases predicted in Ref. \cite{Tong_2_1} are successfully probed by observing dynamical inverse 
participation ratio (IPR) \cite{Yanbo}. 

Decade ago, ME was discovered in a AA-like model with long-range hoppings by Biddle et.al \cite{Sarma_3}. In this paper, they solved the expression of the ME 
by the dual transformation. The exact ME is coincident with the energy spectra embellished by the IPR. In 2015, Ganeshan investigated 
the delocalization-localization properties of an AA-like model with generalized potentials \cite{Sarma_4}. They demonstrated the presence of the MEs in this model 
by means of the dual transformation. The analytical ME accurately partitioned the energy spectrum into the extended and the localized part. For this model, 
Xu et.al. carried out some dynamical investigations \cite{ZhihaoXu}. The dynamical behaviors, such as the wave packet propagation and Loschmidt echo in the intermediate 
regime where mobility edge appears are in contrast to those in the extended and localized regimes. Particularly, Wang et.al. discovered the duality between two typical 
AA-like models \cite{YuchengWang} and invariable MEs are discussed in Ref. \cite{Tong_3}. Besides, MEs are investigated in other quasiperiodic models with 
self-dual symmetry \cite{self_dual_1,self_dual_2,self_dual_3}. 

Non-Hermiticity always brings about some novel quantum phenomena without any analogy to the Hermitian case, such as $\mathcal{PT}$-symmetry breaking \cite{PT_b1,PT_b2,PT_b3}, 
exceptional points \cite{EP_1,EP_2,EP_3,EP_4,EP_5}, anomalous bulk-boundary correspondence \cite{a_bbc}, and non-Hermitian skin effect \cite{skin_1,skin_2,skin_3,skin_4}. 
Furthermore, the interplay between the uncorrelated disorder and  non-Hermiticity will give rise to the delocalization-localization transition \cite{dlt_1,dlt_2,dlt_3,dlt_4,dlt_5,dlt_6,dlt_7,dlt_8}. 
An intriguing discovery is the appearance of MEs in Hanato-Nelson model with nonreciprocal hoppings \cite{HN_1,HN_2,HN_3,HN_4}. A recent study provides an intuitive 
topological explanation why localization transition happens in the Hatano-Nelson model \cite{HN_5}. Not only that, MEs appear in the non-Hermtian systems accompanied 
by the quasi-periodic potentials, showing their robustness against the uncorrelated disorder and non-Hermiticity. For example, Liu et.al. concentrate on the relationship 
between the Anderson localization and the $\mathcal{PT}$-symmetry breaking as well as the MEs in a 1D non-Hermitian quasicrystal \cite{YLiu}. The main findings are 
that Anderson localization is accompanied by the $\mathcal{PT}$-symmetry breaking, and MEs only emerge in the real energy part. Zeng et.al. determined the topological 
nature of MEs in a non-Hermitian AA-like model \cite{Zeng}. Liu et.al. studied two general AA-like models with non-Hermitian potentials, and acquired exact MEs by 
means of the self-dual condition \cite{TLiu}. In experiments, MEs are successfully observed in 3D Anderson models \cite{3D_1,3D_2,3D_3} and 1D AA-like models \cite{1D_1,1D_2,1D_3,1D_4}. 

The delocalization-localization transition and topological superconducting were originally two different research fields, but now are linked by the $p$-wave pairings. 
Kitaev model is a standard superconducting model to interprete the topological transition in topological superconductors \cite{Kitaev}. The AA-like model with $p$-wave 
pairings can be viewed as the quasi-periodic generalizations \cite{Cai} of the Kitaev model, implying that both the two mentioned phenomena are capable of occurring in a 
topological superconductor. Coincidently, the phenomenon that the Anderson localization transition is synchronized with topological superconducting transition is 
uncovered by Cai et.al. \cite{Cai}. Almost at the same time, the transport properties of this quasi-periodic topological superconductor are well discussed \cite{Sen}. 
However, the exploration of the physics behind such types of system is far from over. Recently, the extended-critical are detailedly discussed in the topological non-trivial 
phase with chirally distributed Majorana zero states. Moreover, the extended and critical regions as well as their boundaries completely accord with the predictions 
done by the multifractal analysis \cite{JWang}. In recent years, the relevant studies has extended to other quasi-periodic generations \cite{off_diagonal_pwave_0,off_diagonal_pwave}, 
and besides, the quench dynamics \cite{quench} and Kibble-Zurek machanism \cite{KZ} are well investigated.  

It was studied that the topological non-trivial region of the Kitaev model where Majorana zero mode (MZM) exists is independent of the superconducting pairing strength, 
and is only determined by the hopping amplitude \cite{Kitaev}. The quasi-periodic potential unexpectedly becomes an advantage that it broadens the original topological 
non-trivial region \cite{Cai}, offering multiple degrees of freedom to manipulate the topological superconducting transition. Dramatically, this advantage brought 
by quai-periodicity is greatly fragile in the presence of non-Hermiticity, which will compress the original  non-trivial region \cite{NH_MZM}. But at least, we should reach 
an agreement that MZMs dot not disappear in spite of non-Hermiticity \cite{NH_MZM,NH_MZM_1,NH_MZM_2,NH_MZM_3,NH_MZM_4,NH_MZM_5}.  
In addition, we know that the non-Hermitian systems usually have complex eigenvalues, which are the direct results of nonconservation of probability on account of gain 
and loss. However, Bender and Boettcher found that in a class of systems with $\mathcal{PT}$-symmetry (combination of parity ($\mathcal{P}$) symmetry and 
time-reversal ($\mathcal{T}$) symmetry), there have pure real energies \cite{PT_b1}. The reason why there exists real energies is that $\mathcal{PT}$-symmetry allow 
the gain and loss to be coherently balanced. $\mathcal{PT}$-symmetry breaking indicates that such a balance is broken, then the energies become complex. For decades, 
the $\mathcal{PT}$-symmetry was once regarded as the minimum constraint to preserve the real energies. The latest researches, however, have shattered that perception. 
There exists a type of unconventional real-complex transition independent of $\mathcal{PT}$-symmetry breaking. Hamazaki et.al. found that in a non-Hermitian many-body 
system only with time-reversal symmetry, many-body localization can significantly restrain the imaginary part of complex energies, whereas system with broken time-reversal 
symmetry cannot retain real energies \cite{Hamazaki}. Hereto, does this result mean that the time-reversal symmetry is the lowest constraint to maintain real energies? The 
answer is negative. Reference \cite{NH_MZM} shows that in a class of non-Hermitian topological superconductor only with particle-hole symmetry, the real-complex transition 
still exists. Moreover, different form the consequence of Hamazaki et.al., here the extended phase maintains the real energies. 

 In reality, topological superconductor cannot avoid the exchange of matter and energy with its surroundings, forming the so called non-Hermitian systems,  and this exchange 
 is not conductive to the existence of MZMS \cite{NH_MZM}. Therefore, it is desirable to search a topological superconductor that is robust against the non-Hermitian 
 perturbations. In this paper, we are motivated to theoretically engineer a topological superconductor which is capable of preserving the same topological features as their original Hermitian case, 
 and thus immune to the fragility caused by the non-Hermitian perturbations. A topological phase diagram will be presented by means of the transfer matrix method and the relationship 
 between the topological phase transition and the gap closing will be discussed. Besides, we will analyze its energy spectrum to find if there will be a unconventional real-complex transition 
 independent of $\mathcal{PT}$-symmetry breaking. Furthermore, we try to reveal the correspondence between the real-complex transition and delocalization-localization 
 transition in this system and verify this correspondence by means of the IPR.     
 
 The rest of this paper is organized as follows. In Sec.~\ref{S2}, we describe the non-Hermitian $p$-wave superconductor and present its Hamiltonian both under periodic 
 boundary condition (PBC) and open boundary condition (OBC). In Sec.~\ref{S3}, we introduce the transfer matrix method on the purpose of extracting the $Z_{2}$ topological 
 invariant of this system. In Sec.~\ref{S4}, we investigate the MZMs and obtain the topological phase diagram explicitly, and we explore the existence of the real-complex 
 transition and discuss the correspondence between real-complex transition and the delocalization-localization transition by means of the IPR. A brief summary is given in Sec.~\ref{S5}.

\section{model and Hamiltonian}\label{S2}
We consider a  one-dimensional $p$-wave superconductor with generalized non-Hermitian quasiperiodic on-site potentials, whose Hamiltonian in the real space is 
expressed as

\begin{equation}\label{eq1}
\hat{H}=\sum^{L-1}_{n=1}\left(-t\hat{c}^\dag_{n}\hat{c}_{n+1}+\Delta\hat{c}^\dag_{n+1}\hat{c}^\dag_{n}+h.c.  \right)+\sum^{L}_{n=1}V_{n}\hat{c}^\dag_{n}\hat{c}_{n},
\end{equation}
where $n$ is the site index, $L$ is the size of the system, and $\hat{c}_{n} (\hat{c}^\dag_{n})$ is the fermion annihilation (creation) operator. $t$ in the hopping amplitude 
chosen as the unit of energy, and $\Delta$ denotes the strength of superconducting pairings between nearest-neighbor sites \cite{Kitaev}.  We choose the real $\Delta$ 
which makes it clear that our system do not posses $\mathcal{PT}$-symmetry \cite{PT_b1}, and the system belongs to the class D in the topological classification \cite{Dclass}. 
The generalized non-Hermitian potential $V_{n}$ has the following form
\begin{equation}
V_{n}=\frac{V}{1-be^{i2\pi\alpha n}},
\end{equation}
where $V$ represents the strength of the potential, $b \in (0,1)$ is a dimensionless parameter and $\alpha=(\sqrt{5}-1)/2$ is the incommensurate modulation frequency which 
makes the potential quasiperiodic. Such a complex potential can be experimentally realized in a synthetic mesh photonic lattice and experimental proposal has been discussed in 
Ref.~\cite{TLiu}. When $b=\alpha=0$, the model is reduced to the Kitaev model \cite{Kitaev}, whose topological phase boundary is known at $V=2t$ ($V<2t$ 
is topological non-trivial phase and $V>2t$ is topological trivial phase). 

Due to the particle-hole symmetry, we can make a diagonalization on the Hamiltonian in Eq.~(\ref{eq1}) to obtain its energy spectrum. We perform the Bogoliubov-de Gennes (BdG) 
transformation,
\begin{equation}\label{BdG}
\hat{\xi}^\dag_{j}=\sum^{L}_{n=1}\left[u_{j,n}\hat{c}^\dag_{n}+v_{j,n}\hat{c}_{n}\right],
\end{equation}
where $\hat{\xi}^\dag_{j}$ is the BdG operator, $j$ is the energy level index and it belongs to $1$, $\cdots$, $L$ and the coefficients $u_{j,n}$ and $v_{j,n}$ are complex numbers, 
so that the eigenenergy $E_{j}$ of the system can be determined by 
the following BdG equations
\begin{equation}\label{BdG}
\left\{
\begin{aligned}
-t(u_{n-1}+u_{n+1})+\Delta(v_{n-1}-v_{n+1})+V_{n}u_{n}&=E_{j}u_{n},\\
t(v_{n-1}+v_{n+1})+\Delta(u_{n+1}-u_{n-1})-V_{n}v_{n}&=E_{j}v_{n}.
\end{aligned}
\right.
\end{equation}
To make the above equation a matrix presentation, we introduce the wave function with the following form
\begin{equation}
\ket{\psi_{j}}=(u_{j,1},v_{j,1},u_{j,2},v_{j,2},\cdots,u_{j,L},v_{j,L})^{T}.
\end{equation}
At $n$-th lattice site, the probability $p_{n}$ is given as $p_{n}=u^2_{n}+v^2_{n}$. Then, we arrive at the BdG matrix
\begin{equation}\label{matrix}
\mathcal{H}=\left(
\begin{array}{ccccccc} 
A_{1} & B & 0 & \cdots & \cdots & \cdots & C \\
B^\dag & A_{2} & B & 0 & \cdots & \cdots & 0 \\
0 & B^\dag & A_{3} & B & 0 & \cdots & 0 \\
\vdots & \ddots & \ddots & \ddots & \ddots & \ddots & \vdots \\
0 &\cdots & 0 & B^\dag & A_{L-2} & B & 0\\
0 &\cdots & \cdots & 0 & B^\dag & A_{L-1} & B \\
C^{\dag} & \cdots & \cdots & \cdots & 0 & B^\dag & A_{L} 
\end{array}
\right), 
\end{equation}
where 
\begin{equation}
A_{j}=\left(
\begin{array}{cc}
V_{j} & 0 \\
0 & -V_{j}
\end{array}
\right), 
B=\left(
\begin{array}{cc}
-t & -\Delta\\
\Delta & t
\end{array}
\right), 
\end{equation}
 and 
 $C=\left(
\begin{array}{cc}
-t & \Delta\\
-\Delta & t
\end{array}
\right)$ for system with periodic boundary condition (PBC) and $C=\left(
\begin{array}{cc}
0 & 0\\
0 & 0
\end{array}
\right)$ for system with open boundary condition (OBC). 
 
Definitely, $\mathcal{H}$ is a $2L \times 2L$ matrix. By diagonalizing $\mathcal{H}$, we can acquire the full energy spectrum $E_{j}$ 
and the corresponding wave functions $\ket{\psi_{j}}$ directly. 

In the next section, we will discuss the topological properties of the system, such as the $Z_{2}$ topological invariant, Majorana zero energy modes and the corresponding states. 
Moreover, we will quantitatively analyze real-complex energy transition and the mobility edge by the inverse participation ratio.

\section{Transfer Matrix method}\label{S3}
As mentioned that our system belongs to the class D in the topological classification, the topological properties of the system is directly reflected by a 
$Z_{2}$ topological invariant $M$. We determine $M$ by the scattering matrix method \cite{scatter_1,scatter_2}, for the reason that this method is well-behaved 
in non-Hermitian topological superconductor \cite{NH_MZM}. The scattering matrix $\bm{S}$ consists of four submatrices with the 
following form 
\begin{equation}
\bm{S}=\left(
\begin{array}{cc}
\bm{R} & \bm{T'}\\
\bm{T} & \bm{R'}
\end{array}
\right),
\end{equation}
in which the $2\times 2$ subblocks $\{\bm{R}, \bm{R}'\}$ and $\{\bm{T}, \bm{T}'\}$ denotes the reflection and transmission matrices at the two ends of the 
system. With the matrix $\bm{R}$, the $Z_{2}$ topological invariant $M$ is given as 
\begin{equation}
M=sgn(\rm{Det}(\pmb{R})), 
\end{equation}
where $sgn$ is the sign of the determinant (\rm{Det}) of $\bm{R}$. When $M=-1$, the system is in the topological nontrivial phase, which supports the 
existence of the Majorana zero-energy mode (MZM), and $M=1$ corresponds to the topological trivial phase where there is no any MZM. 

The scattering matrix $\bm{S}$ can be acquired by the transfer matrix scheme. We set the Fermi energy at $E_{f}=0$, then Eq.~(\ref{BdG}) with zero energy 
is rewritten as 
\begin{equation}
\left(
\begin{array}{c}
\hat{t}^\dag_{n}\bm{\phi}_{n}\\
\bm{\phi}_{n+1}
\end{array}
\right)=\bm{\tilde{\lambda}}_{n}\left(
\begin{array}{c}
\hat{t}^\dag_{n-1}\bm{\phi}_{n-1}\\
\bm{\phi}_{n}
\end{array}
\right),
\end{equation}
in which $\bm{\phi}_{n}=(u_{n}, v_{n})^{\rm{T}}$ denotes the wave function at $n$-th site, and 
\begin{equation}
\bm{\tilde{\lambda}}_{n}=\left(
\begin{array}{cc}
0 & \hat{t}^\dag_{n}\\
-\hat{t}^{-1}_{n} & -\hat{t}^{-1}_{n}\hat{\tau}_{n}
\end{array}
\right)
\end{equation}
with $\hat{t}_{n}=-t\bm{\sigma_{z}}+i\Delta\bm{\sigma}_{z}$ and $\hat{\tau}_{n}=V_{n}\bm{\sigma_{z}}$. After a recursive 
process, the waves at the two ends ($n=1$ and $n=L$) of the system depend on the total transfer matrix $\bm{\tilde{\lambda}}$  
\begin{equation}
\bm{\tilde{\lambda}}=\bm{\tilde{\lambda}}_{L}\bm{\tilde{\lambda}}_{L-1} \cdots \bm{\tilde{\lambda}}_{2}\bm{\tilde{\lambda}}_{1}.
\end{equation}
In order to separate the left-moving and right-moving waves, we need to introduce a unitary operators $\bm{U}$. After performing 
a unitary transformation, the total transfer matrix becomes 
\begin{equation}
\bm{\lambda}=\bm{U^{\dag}}\bm{\tilde{\lambda}}\bm{U}, 
\bm{U}=\frac{1}{\sqrt{2}}\left(
\begin{array}{cc}
\bm{I} & \bm{I} \\
i\bm{I} & -i\bm{I}
\end{array}
\right),
\end{equation}
where $\bm{I}$ is a identity matrix. Accordingly, the relationship between $\bm{R}$ ($\bm{R}'$) and $\bm{T}$ ($\bm{T}'$) 
is presented as 
\begin{equation}
\left(
\begin{array}{c}
\bm{T}\\
\bm{0}
\end{array}
\right)=\bm{\lambda}
\left(
\begin{array}{c}
\bm{I}\\
\bm{R}
\end{array}
\right), 
\left(
\begin{array}{c}
\bm{R}'\\
\bm{I}
\end{array}
\right)=\bm{\lambda}
\left(
\begin{array}{c}
\bm{0}\\
\bm{T}'
\end{array}
\right). 
\end{equation}
Therefore, we can calculate the total transfer matrix $\bm{\lambda}$ to obtain the $Z_{2}$ topological invariant $M$. 

\begin{figure}[t]
\centering
\includegraphics[width=0.5\textwidth]{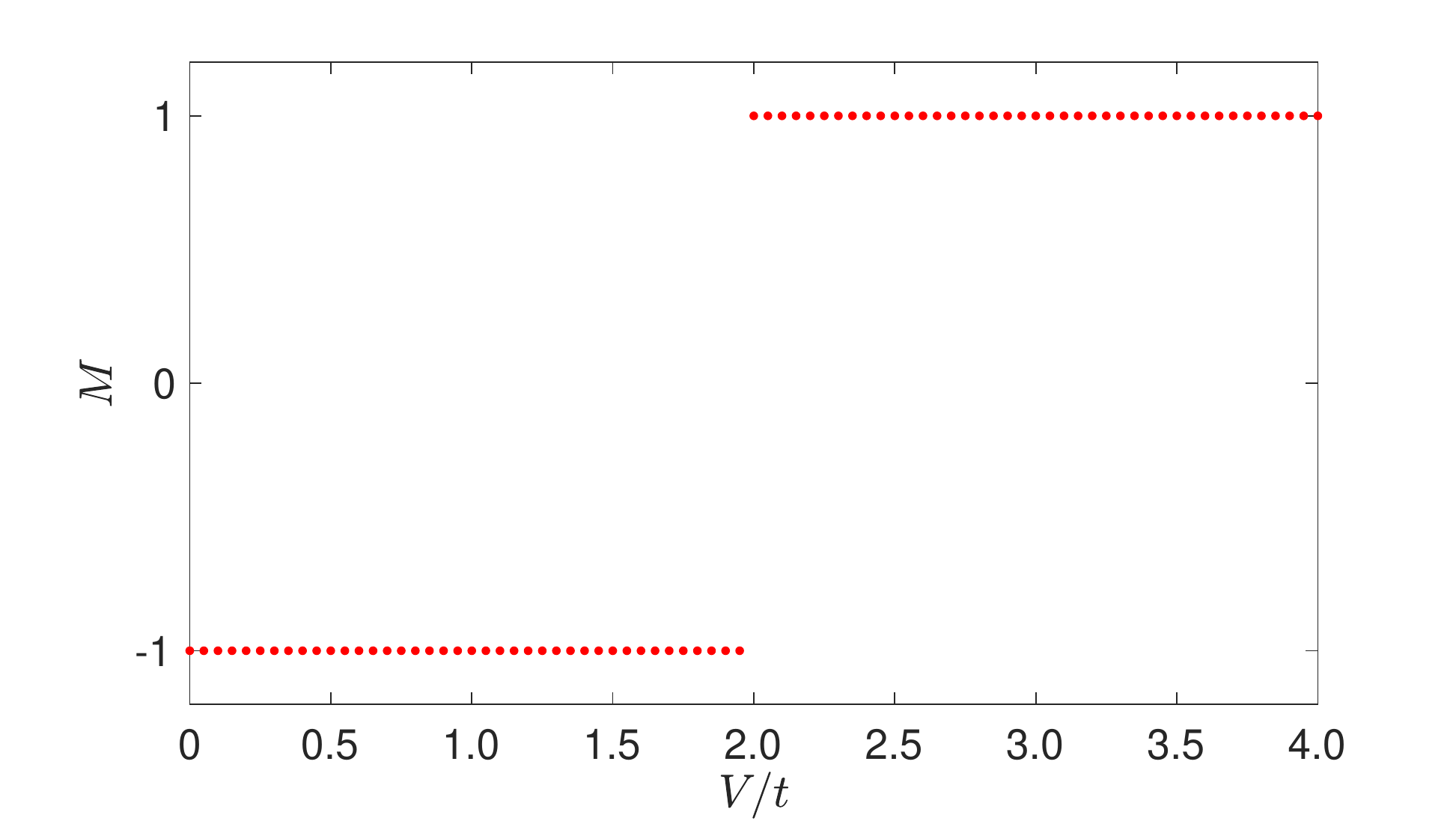}\\
\caption{(Color Online) $Z_{2}$ topological invariant $M$ as a function of the potential strength $V$.  $M=-1$ refers to the topological non-trivial phase, 
and $M=1$ shows the topological trivial phase. Other parameters are $t=1$, $\alpha=(\sqrt{5}-1)/2$ and $\Delta=0.5t$.}\label{f1}
\end{figure}

\section{results and discussions}\label{S4}

At the beginning, we investigate the topological properties of this system. For convenience but not losing generality, we take t=1 and $\Delta=0.5$ 
throughout the following studies. Along the above strategy, we numerically calculate the $Z_{2}$ topological invariant $M$ with various potential 
strength $V$ and finally acquire the phase diagram, which is explicitly shown in Fig.~\ref{f1}. Intuitively, there are two different phases with $M=-1$ 
and $M=1$, respectively. According to Refs.~\cite{NH_MZM}, we know that here $M=-1$ stands for the topological non-trivial phase and $M=1$ denotes 
the topological trivial phase. For $b=0$, we have already known that our model is reduced to the Kitaev model, whose topological boundary is $V=2t$. 
For other different $b$, the topological boundary is also stably located at about $V=2t$, reflecting that the topological properties are immune to the 
non-Hermtian disturbance. This feature is of importance to its practical applications. Moreover, we note that the topological boundary is the same as 
that of Kitaev model. From this aspect, our theoretical model is the reappearance of the Kitaev model in the non-Hermtian case.   

\begin{figure}[H]
\centering
\includegraphics[width=0.5\textwidth]{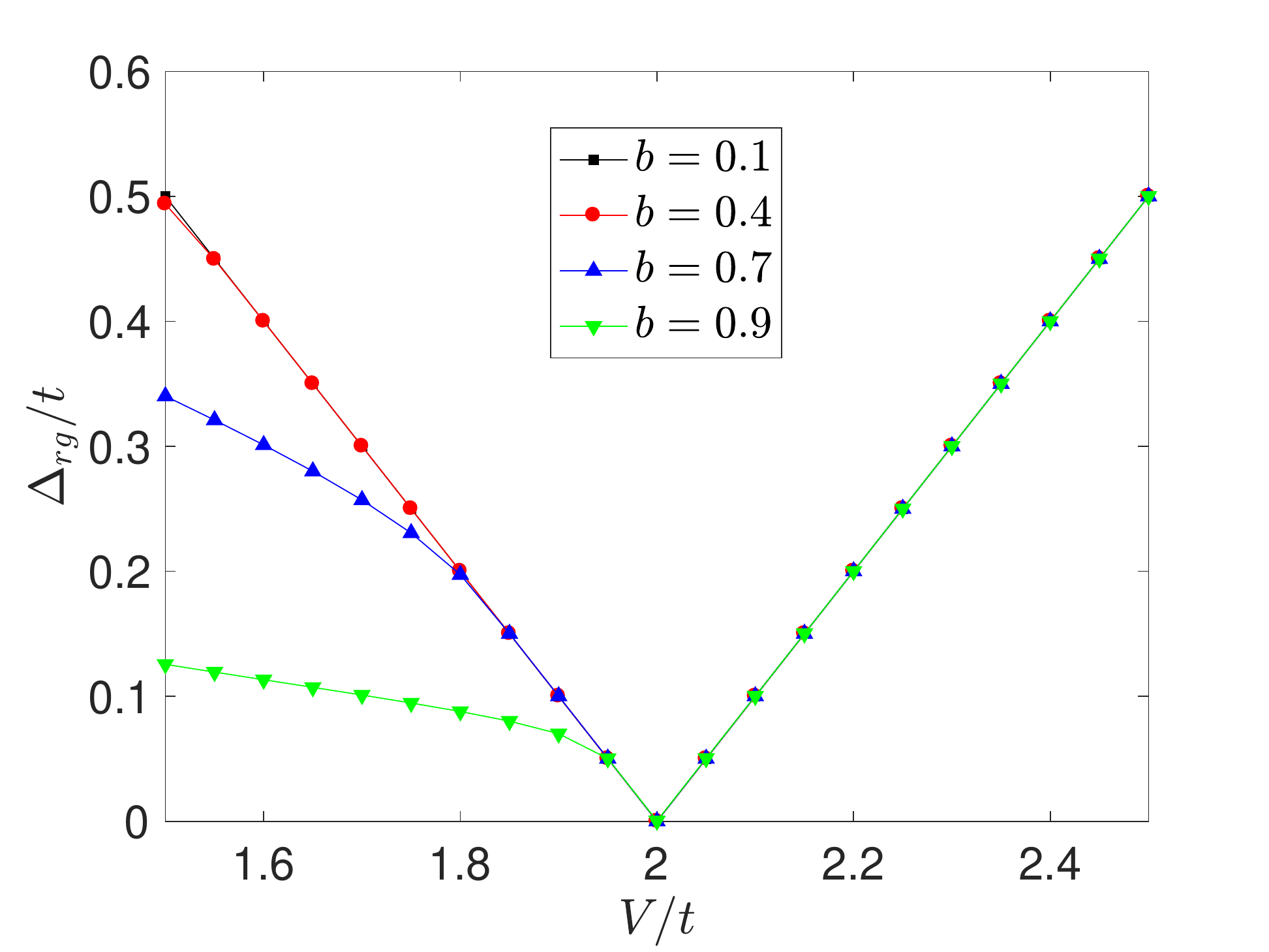}\\
\caption{(Color Online) The variation of real energy gap $\Delta_{rg}$ as a function of  $V$ with various $b$. Other involved parameters are $\alpha=(\sqrt{5}-1)/2$, 
$\Delta=0.5t$, and $L=500$.}\label{f2}
\end{figure}

For topological insulators and Chern insulators, gap closing is a key feature to manifest the topological phase transition. We find that this characteristic is 
not an exception in our non-Hermitian $p$-wave superconductor. With the purpose of making this conception explicit, we investigate that how the real gap 
$\Delta_{rg}$ behaves with potential strength $V$. $\Delta_{rg}$ is the difference of the ($L+1$)-th real energy level and the $L$-th real energy level 
under PBC with the definition as $\Delta_{rg}=E_{L+1}-E_{L}$. We choose four typical values of parameter $b$ ($b=0.1$, $0.4$, $0.7$, and $0.9$) and fix the 
size of the system $L=500$, then the corresponding energies can be extracted by diagonalizing the matrix presented in Eq.~(\ref{matrix}). Figure \ref{f2} shows 
the variation of $\Delta_{rg}$ as a function of the potential strength $V$ with various $b$. For different $b$, the real gap of this system is readily seen to be 
closed at $V=2t$. What needs illustration is that for larger size of the system, gap closing point still locates at $V=2t$.  This result is in accordance 
with the phase diagram in Fig.~\ref{f1}, and confirms the prediction that topological superconducting transition is accompanied by the gap closing.

\begin{figure}[H]
\centering
\includegraphics[width=0.5\textwidth]{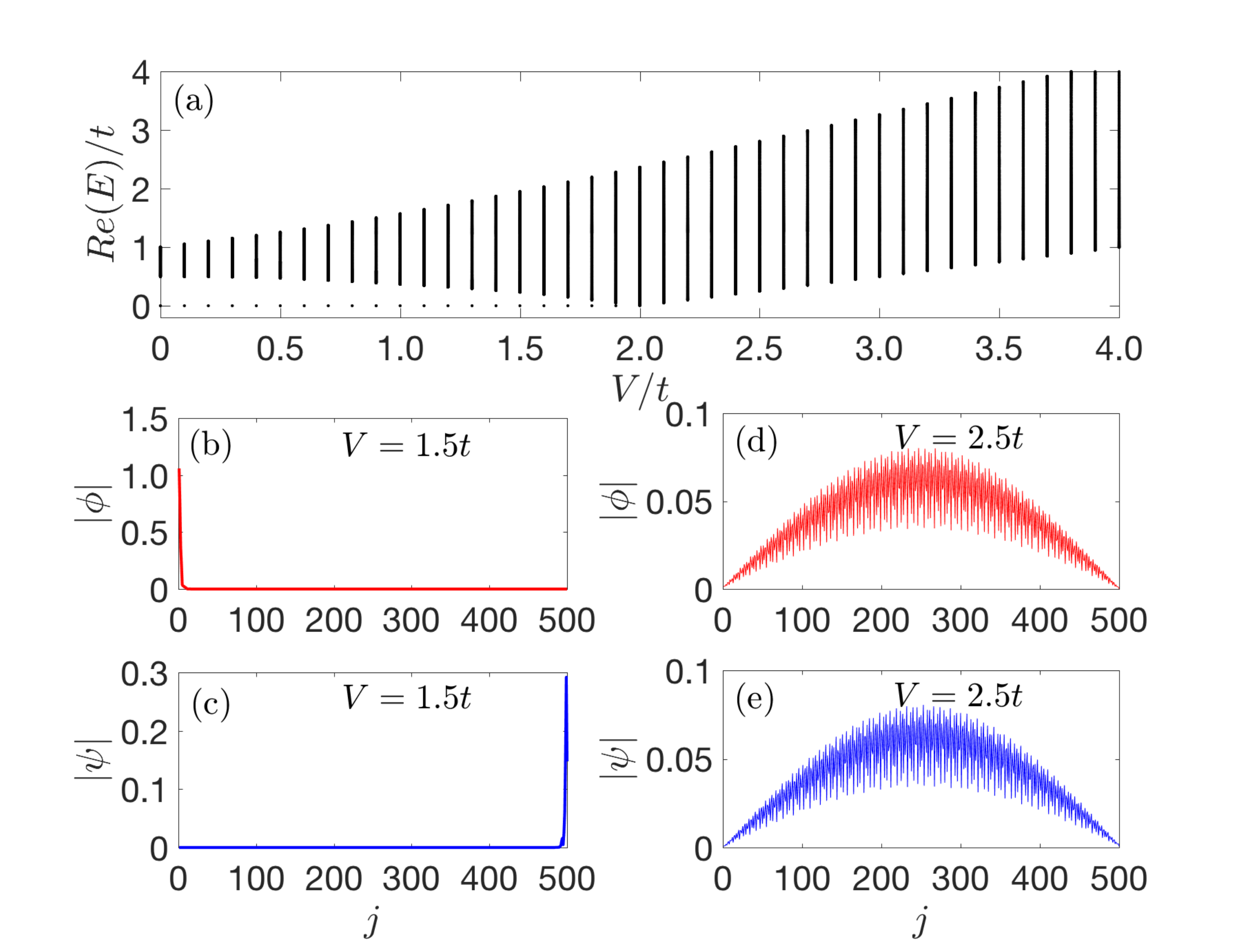}\\
\caption{(Color Online) Top panel:  (a) The real part of excitation spectrum under OBC. Bottom panel: Spatial distributions of $|\phi|$ and $|\psi|$ for 
the lowest excitations with $V=1.5t$ in (b) and (c) respectively and with $V=2.5t$ in (d) and (e) respectively. $V=2t$ is the topological phase transition 
point of the system. Other involved parameters are $b=0.5$, $\alpha=(\sqrt{5}-1)/2$, and $L=500$.}\label{f3}
\end{figure}

Now that we have figured out the topological phases of the system, what other physical information can be extracted from the topological non-trivial phase? 
By considering OBC, $b=0.5$, and $L=500$, we acquire the real part of excited spectrum of the superconductor, shown in Fig.~\ref{f3}(a). As the picture shows, there are 
MZMs in the topological non-trivial phase ($V<2t$), while the MZM in the trivial phase ($V>2t$) is absent. In other words, MZM is protected by the topology. 
We further want to investigate the bulk-edge correspondence by analyzing the lowest excitation mode. With this purpose, we rewrite the BdG operator in Eq.~(\ref{BdG}) as 
\begin{equation}
\eta^\dag_{j}=\frac{1}{2}\sum^{L}_{n=1}[\phi_{j,n}\gamma^{A}_{n}-i\psi_{j,n}\gamma^{B}_{n}],
\end{equation}
where $\gamma^{A}$ and $\gamma^{B}$ are Majorana operators, satisfying $\gamma^{A}=\hat{c}^\dag_{n}+\hat{c}_{n}$ and $\gamma^{A}=i(\hat{c}^\dag_{n}-\hat{c}_{n})$ 
and obeying the relations $(\gamma_{n}^\beta)^\dag=\gamma_{n}^\beta$ and $\{\gamma_{n}^{\beta},~\gamma_{n'}^{\beta'}\}=2\delta_{nn'}\delta_{\beta\beta'}$ 
with $\beta, \beta'$ $\in \{A, B\}$;  $\phi_{j,n}=(u_{j,n}+v_{j,n})$ and $\psi_{j,n}=(u_{j,n}-v_{j,n})$. 

Figures \ref{f3}(b) and \ref{f3}(d) (Figures \ref{f3}(c) and \ref{f3}(e)) present spatial distributions of $|\phi|$ ($|\psi|$) for their corresponding lowest excitation modes. 
According to the real excitation spectrum, we immediately know that when $V=1.5t$, the lowest excitation mode is exactly the MZM.  It can be seen that $|\phi|$ and $|\psi|$ 
are distributed at the disparate ends of the superconductor, implying that original paired Majorana fermions (MFs) have been split into two independent and spatial-separated 
MFs. This phenomenon is the direct result of non-trivial topology. Moreover, no anomalous edge state \cite{a_bbc} is found here. In the same way, we know that when $V=2.5t$, the 
lowest excitation mode ceases to be the MZM, but corresponds to the bulk state. As the Figs.~\ref{f3}(d) and \ref{f3}(e) show, $|\phi|$ and $|\psi|$ are distributed in the bulk 
of the system, signifying that there is no spatial-separated MF. This circumstance is determined by the trivial topology.

\begin{figure}[H]
\centering
\includegraphics[width=0.5\textwidth]{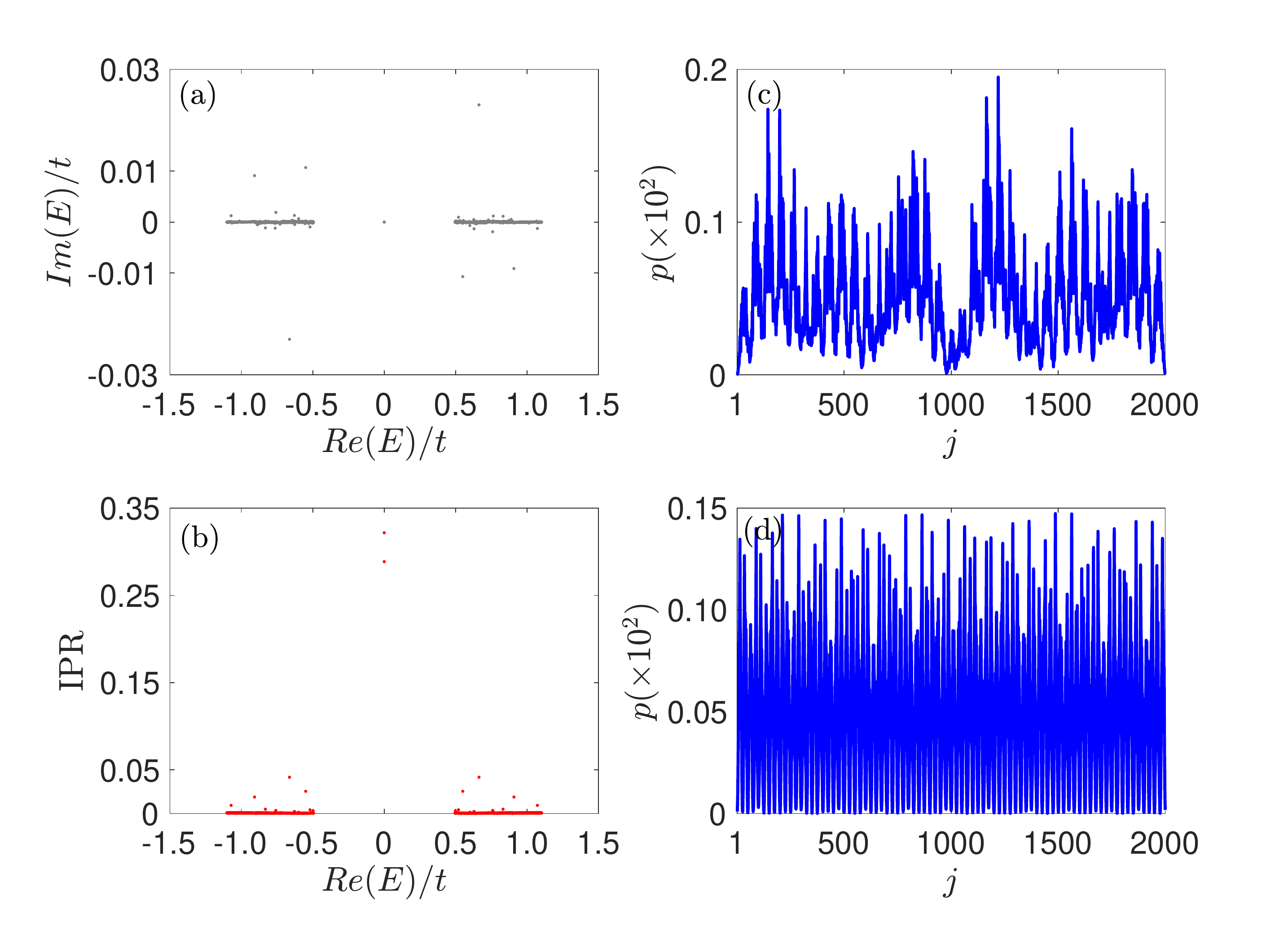}\\
\caption{(Color Online) (a) Energy spectrum in the complex plane with $V=0.2t$. (b) IPR versus $Re(E)$. (c) and (d) are typical extended wave functions 
taken from the $71$-th and $1921$-th excitation modes, respectively. Other involved parameters are $\alpha=(\sqrt{5}-1)/2$ and $L=2000$. }\label{f4}
\end{figure}

For non-Hermitian systems, we intuitively take the attitude that their eigenvalues are complex. Nevertheless, the theorem proposed by Bender and Boettcher 
points out that there are real energies in systems with $\mathcal{PT}$-symmetry \cite{PT_b1}. When this symmetry is broken, then the energies become complex. Recently, 
there was a research that discovered an unconventional real-complex energy transition independent of $\mathcal{PT}$-symmetry breaking \cite{NH_MZM}. This 
intriguing finding motivates us to make it clear if there is other types of real-complex transition that is not dominated by $\mathcal{PT}$-symmetry breaking. For the sake of general 
consideration, we take $b=0.5$, $L=2000$, and OBC in the following analysis. Figure \ref{f4}(a) plots the energy spectrum in the complex plane with $V=0.2t$. System 
now is in the topological trivial phase.  Intuitively, the imaginary part of energies are suppressed at $Im(E)=0$. We conjecture that the real energies are closely related 
to the extended states. To validate this speculation, we introduce the IPR: 
\begin{equation}
{\rm{IPR}}_{j}=\sum^{L}_{n=1}\left(|u_{j,n}|^4+|v_{j,n}|^4\right).
\end{equation}
For a normalized wave function, if its IPR tends to zero, then this wave function is extended; if IPR is greater than zero (approaching $1$), then 
this wave function is localized. Figure \ref{f4}(b) shows the variations of IPR with $Re(E)$ at $V=0.2t$. It is readily seen that IPRs are equal to zero, 
implying all these wave functions are extended. Figures \ref{f4}(c) and \ref{f4}(d) are probability distributions of the typical extended wave functions 
taken from the $71$-th and $1921$-th excitation modes, respectively. Readily, wave functions 
extends throughout the system, conform to the analysis by the IPR. Here, $p$ denotes the probability, satisfying $p_{n}=u^2_{n}+v^2_{n}$, and the same below. 

\begin{figure}[H]
\centering
\includegraphics[width=0.5\textwidth]{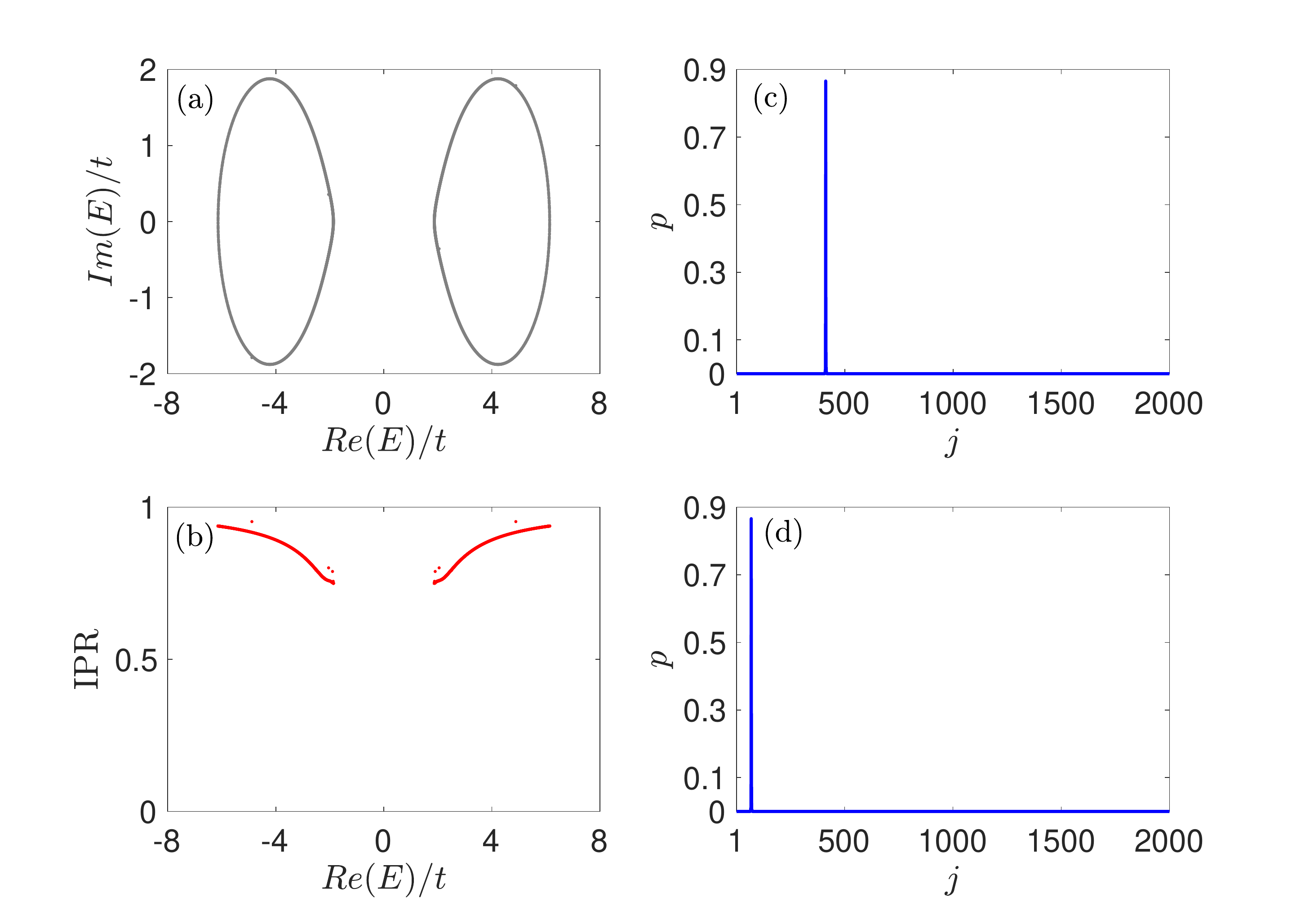}\\
\caption{(Color Online) (a) Energy spectrum in the complex plane with $V=6t$. (b) IPR versus $Re(E)$. (c) and (d) are possibility distributions of typical 
localized wave functions taken from the $520$-th and $578$-th excitation modes, respectively. 
Other involved parameters are $\alpha=(\sqrt{5}-1)/2$ and $L=2000$. }\label{f5}
\end{figure}

If we take $V=6t$ and the system is in the topological trivial phase, then we will observe distinctly different phenomena. Compared to the energy spectrum in 
Fig.~\ref{f4}(a), the energies at $V=6t$ shown in Fig.~\ref{f5}(a) are fully complex. The spectrum presents a closed loop, indicating that there 
is no skin effect \cite{skin_3,skin_4}. We plot the IPR as a function of $Re(E)$ in Fig. ~\ref{f5}(b). As the figure shows, IPRs are finite numbers, approaching $1$, 
which indicate that these wave functions are localized. We choose two typical localized wave functions as two intuitive examples. Figs.~\ref{f5}(c) (the $520$-th 
excitation mode) and \ref{f5}(d) (the $578$-th excitation mode) show that the wave functions are localized in the bulk of the system. 

 \begin{figure}[H]
\centering
\includegraphics[width=0.5\textwidth]{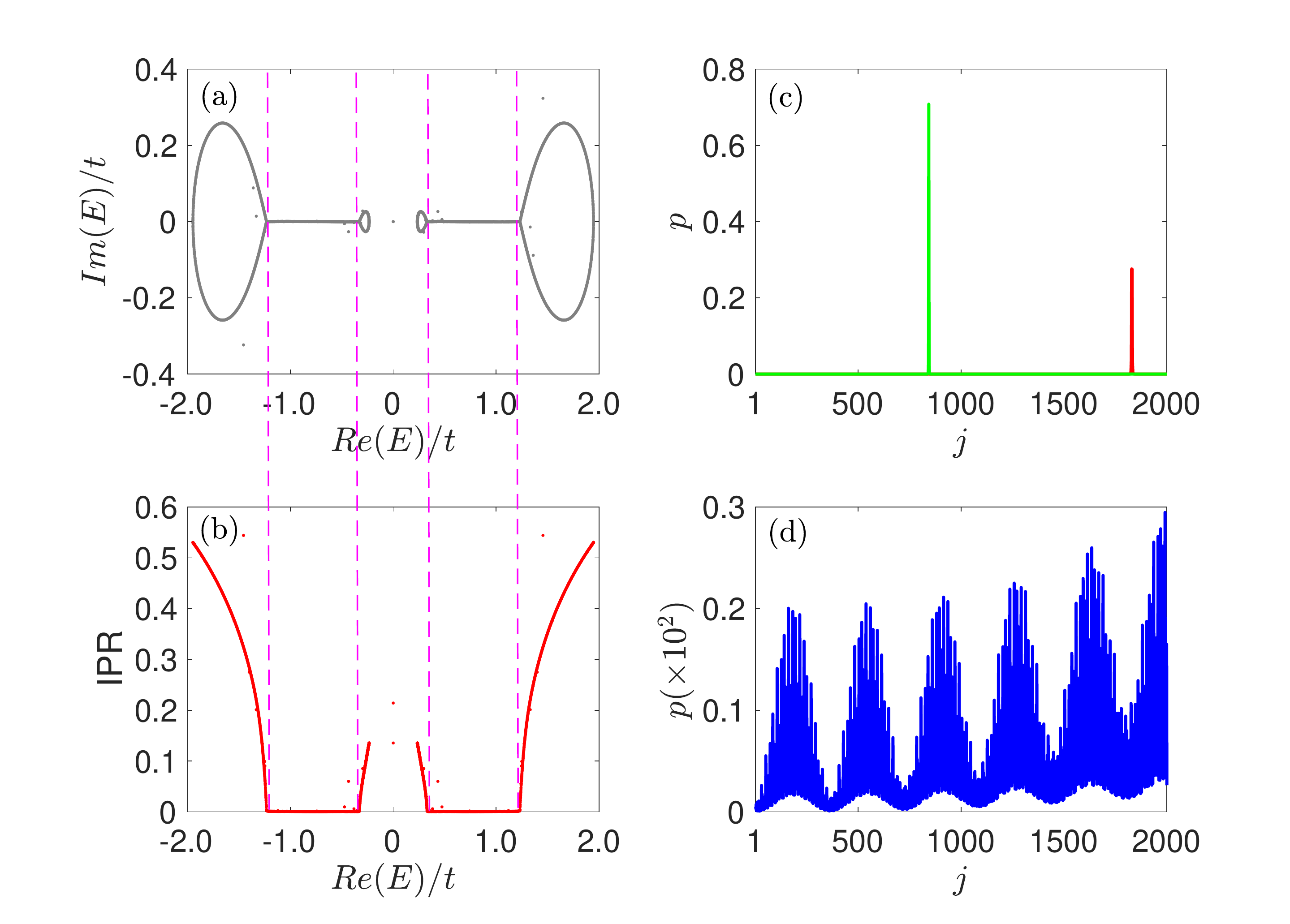}\\
\caption{(Color Online) (a) Energy spectrum in the complex plane with $V=1.5t$. (b) IPR versus $Re(E)$. (c) Red line denotes the probability distribution of a 
typical localized wave function taken from the $10$-th excitation mode whose corresponding energy is located at the small energy loop and green 
denotes the one taken from the $1990$-th excitation mode whose corresponding energy is located at the big energy loop. (d) Probability 
distribution of a typical extended wave function taken from the $578$-th excitation mode with pure real energy. Other involved parameters are $\alpha=(\sqrt{5}-1)/2$ 
and $L=2000$. }\label{f6}
\end{figure}

 From the above numerical analysis, we construct such a correspondence that real energies correspond to the extended states and complex energies to the 
 localized ones. Motivated by this correspondence, we wander to know that whether there exists another type of real-complex transition, i.e., when given 
 parameters, the energies consist of pure real part and complex part. The reason why we pay attention to this issue is that this case will lead to the appearance 
 of MEs. To reveal its existence, we take $V=1.5t$. As discussed before (see the real excitation spectrum in Fig.~\ref{f3}(a)), at $V=1.5t$ the systems is obviously 
 in the topological non-trivial phase and possess the MZM. However, in this case, the spectrum in the complex plane presents different features.

Concrete speaking, the spectrum consists of two small loops and two big loop, as well as two regions of real energy (see Fig.~\ref{f6}(a)). Besides, the IPR of 
loop regions are finite numbers, larger than zero, whereas the ones of real-energy regions are zero (see Fig.~\ref{f6}(b)). In Fig.~\ref{f6}(c), the red line shows 
the probability distribution of a typical localized wave function taken from the $10$-th excitation mode whose corresponding energy is located at the small energy loop 
and the green line denotes the one taken from the $1990$-th excitation mode whose corresponding energy is located at the big energy loop. Differently, 
in Fig.~\ref{f6}(d), we see a extended probability distribution whose corresponding wave function is taken from the $578$-th excitation mode with pure real energy. 
The difference of the characteristics of wave functions announces that the presence of the ME. Intuitively seen from Fig.~\ref{f6}(a), these two 
types of correspondences are separated by four magenta dashed lines, i.e., the so called MEs.

 \begin{figure}[H]
\centering
\includegraphics[width=0.5\textwidth]{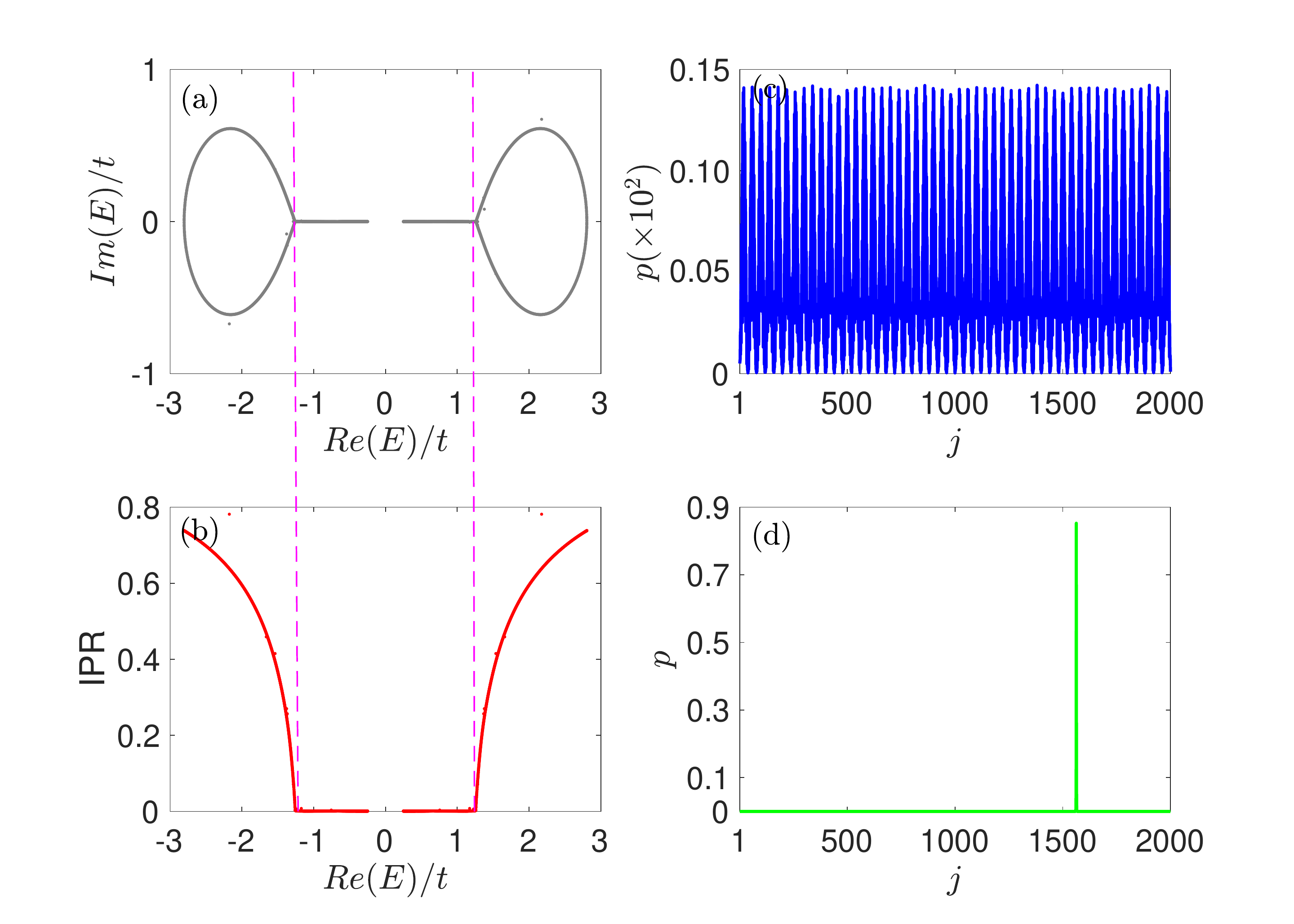}\\
\caption{(Color Online) (a) Energy spectrum in the complex plane with $V=2.5t$. (b) IPR versus $Re(E)$. (c) Possibility distribution of a typical extended 
wave function taken from the $50$-th excitation mode with pure real energy.  (d) Possibility distribution of a typical localized wave function 
taken from the $1950$-th excitation mode with complex energy. Other involved parameters are $\alpha=(\sqrt{5}-1)/2$ and $L=2000$. }\label{f7}
\end{figure}

Similarly, in the same nontrivial topological phase, the properties of energy spectrum and delocalization-localization transition will also be different. As the Fig.~\ref{f7} 
shows, the spectrum at $V=2.5t$ consists of two loop regions and two real energy regions. The transition of energy from complex to real appropriately describes 
the transition of the wave functions from the localized to extended ones by the IPR. The separation lines (magenta dashed lines) are numerical MEs. As two intuitively 
physical pictures, Fig~\ref{f7}(c) shows the probability distribution of a typical extended wave function taken from the $50$-th excitation mode with pure real energy and 
Fig.~\ref{f7}(d) shows the probability distribution of a typical localized wave function taken from the $1950$-th excitation mode with complex energy.

\section{summary}\label{S5}
All in all, we have investigated the topological properties, energy spectrum features, and delocalization-localization properties of a general non-Hermitian $p$-wave superconductor. 
We have known that the system is robust against the non-Hermitian perturbations and preserves the same topological boundary as the original Kitaev model. By calculating the IPR, we 
have successfully contributed a general correspondence that real energies correspond to the extended states and the complex ones correspond to the localized states. This 
correspondence will be instructive to detect the unconventional real-complex transition and delocalization-localization transition in other non-Hermitian systems. Our work directly 
connects three hot research areas: topological superconductors, real-complex transition induced by non-Hermiticity, and Anderson localization, and will promote comprehensive 
development of these areas. Unfortunately, it is unknown whether there exists a type of symmetry breaking  which results in this unconventional real-complex transition. Therefore, 
revealing the origination of this unconventional real-complex transition  remains an open question.

\section*{acknowledge}
We acknowledge the support from NSFC under Grants No.11835011 and No.11774316.


\end{document}